\documentclass[a4paper,10pt]{article}

\usepackage{makeidx}               
\usepackage{graphicx}              
\usepackage[bookmarksnumbered,colorlinks,plainpages,citecolor=blue,urlcolor=blue]{hyperref}
\usepackage[rflt]{floatflt}
\usepackage{subfigure}
\usepackage{caption2}
\usepackage{latexsym}
\usepackage{amsmath}
\usepackage{amsfonts}
\usepackage{amssymb}
\usepackage{enumerate}

\newcommand{\uck}[1]{\o}

\newcommand{\ket}[1]{\mbox{$|#1\protect\rangle$}}

\def\beq{\begin{equation}}
\def\eeq{\end{equation}}
\def\bea{\begin{eqnarray}}
\def\eea{\end{eqnarray}}

\begin{document}

\title{ Communication Complexity Protocols for Q-trits}

\vspace{15 mm}
\author{Boaz Tamir$^a$\footnote{\texttt{tamirb@popeye.os.biu.ac.il }} \\
\and}

\date{}

\maketitle
$^a$ Department of philosophy of science, \\
Bar-Ilan University, Ramat-Gan, Israel

\begin{abstract}
\noindent
\hspace{5mm}

Consider a function where its entries are distributed among many parties. Suppose each party is allowed to transmit only a limited amount of information to a net. One can use a classical protocol to guess the value of the global function. Is there a quantum protocol improving the results of all classical protocols? In \cite{Brukner2004} Brukner et al. showed the deep connection between such problems and the theory of Bell's inequalities. Here we generalize the theory to trits. There, the best classical protocol fails whereas the quantum protocol yields the correct answer.

\end{abstract}
\section{Introduction.}

\hspace{5mm}Suppose we want to compute some global function where the entries are distributed among several parties. Suppose also we have a limited number of bits each party can communicate to a net. Can we compute the global function? How far can we guess its values using a classical protocol? Is there a quantum protocol improving our guess?

In \cite{Buhrman97} Buhrman et al. showed such a communication complexity problem for two parties. The probability of success for the best classical protocol was 0.75, whereas a quantum protocol was good within an accuracy of 0.86.

In \cite{Brukner2004} Brukner et al. showed the deep connection between a class of communication complexity problems (CCP) and the theory of Bell inequalities. To each such CCP one can attach a Bell inequality. Therefore the fact that there exists a quantum protocol improving the probability of success indicates the existence of a quantum state violating this inequality. As a corollary, the theory in \cite{Buhrman97} was extended to the case of many parties. There for a large enough number of participants the probability of success of the classical protocol was very close to 0.5, whereas the quantum protocol always yielded the correct answer.

Here we generalize the above results to trits. A two party protocol for trits was originaly suggested in \cite{Brukner2002}. Here, the entries to the global function are trits and bits that are distributed among many distant parties. The global function can have three values. Then, for a large enough number of participants the probability of success for the best classical protocol is close to $\frac{1}{3}$, that is, the classical protocols fails completely, whereas the quantum protocol yields the correct answer. 

The protocol suggested here can probably be generalized to the case where the global function can have more than three values. 

\section{Communicating with one trit and an entanglement.}

\hspace{5mm}Suppose $k=3n+1$ $(n\geq3)$ parties communicate to compute a global function. Suppose each party holds a register with one trit $Y^i$ and one bit $X^i$  ($i=1,...,3n+1$). Denote the (column) vector of $Y^i$'s (res. $X^i$'s) by ${\textbf{Y}}$ (res. ${\textbf{X}}$). Suppose also exactly $m$ of the $X^i$'s are zero for some $m\in\{0,3,6,...,3n\}$. Define $l({\textbf{X}})=\frac{m}{3}$  $mod(3)$. Let the global function be defined as:

\[ G({\textbf{Y}} ,{\textbf{X}} )=\{\sum_{i=1}^{3n+1}Y^i\} + l({\textbf{X}})\hspace{4mm}  mod(3) \]

The parties do not know $l$ in advance. 

Each one of the players can transmit only one trit of information. We now let the parties share a certain entanglement. The parties will 'write' information on the entanglement using local unitary transformations and according to the value of his/her bit $X^i$. Then each of them will measure the entanglement and add the result $x^i$ to the value of the trit $Y^i$  he/she holds.

We now describe the entanglement and the local operators. 

\textbf{The entanglement:}

Let $[j]^{i}$, $i>0$ denote the entanglement of all $i$-tensors such that their sum of entries is j modulo 3. For example

\[ [1]^3 = |001> + |010> + |100> + |211> + |121> 
+ |112> + |220> + |202> + |022>. \]
 
The entanglement used here is: $[0]^k$

Before presenting the protocol we need a Lemma.

\textbf{Lemma:}

Consider the following permutation $P$ of elements: $|0>$ goes to $|1>$, $|1>$ goes to $|2>$, and $|2>$ goes to $|0>$. Then   

\[ P = \left(\begin{array}{ccc} 0&0&1\\ 1&0&0\\ 0&1&0\end{array}\right)\]

We write $P$ in its eigenvalue basis. Then:

\[ P = S^{-1} A  S \]

where \[ S = \left(\begin{array}{ccc} 1&1&1\\ 1&a&a^2\\ 1&a^2&a\end{array}\right)\]

and \[ A =\left(\begin{array}{ccc} 1&0&0\\ 0&a&0\\0&0&a^2\end{array}\right) \]

\noindent where $a$ is a third root of unity. Compute now a third root of $P$.

\[ P^{1/3} = S^{-1} A^{1/3} S. \]

Then $P^{1/3} \otimes P^{1/3}\otimes P^{1/3}$ operates on 3-tensors and takes $[0]^3$ into $[1]^3$, $[1]^3$ into $[2]^3$, and $[2]^3$ into $[0]^3$.

\textbf{Proof:} Straightforward.

\textbf{Example:} For dimension 2 the above Lemma states that the unitary operator:

\[\frac{1}{2}\left( \begin{array}{cc} 1+i&1-i\\ 1-i&1+i \end{array}\right)\]

\noindent swaps $\ket{01}+\ket{10}$ with $\ket{00}+\ket{11}$

\textbf{The protocol:}

Operate by $P^{1/3}$ whenever the value of the bit $X^i$ is 0 (otherwise apply the identity operator).

\textbf{Lemma:} The above protocol takes $[0]^k$ to $[l]^k$

\textbf{Proof:} For $m\in \{3,6...\}$ decompose $[0]^k$ as follows:

\[ [0]^k= [0]^m[0]^{k-m} + [1]^m [2]^{k-m} + [2]^m [1]^{k-m}\]

Note that the $m-$tensors at the decomposition are the tensors of exactly the places where the parties hold registers with bit equals 0. Such a decomposition always exists.

Now write $[p]^m$ for p=0,1,2 as

\[ [p]^m=\sum_{i_1+...i_j=p \hspace{2mm} mod(3)}[i_1]^3 \cdot \cdot \cdot [i_j]^3\]

\noindent where $j=\frac{m}{3}$. The Lemma now follows from the previous Lemma.

Now each player measures the entanglement and adds the result $x^i$ (which is a trit now) to his/her trit $Y^i$. Therefore

\[ \sum (Y^i+x^i)= \sum Y^i+\sum x^i= {\sum Y^i} +l= G({\textbf{Y}},{\textbf{X}})   \]

Note that each player transmits only one trit $Y^i+x^i$.

\section{ The classical scheme }  

\hspace{5mm} Consider a nondeterministic protocol \cite{Kushilevitz1996} where each one of the players is allowed to send only one trit. Suppose also they all send their trits simultaneously. Each player divides the set
$\{(0,0), (0,1), (1,0), (1,1), (2,0), (2,1)\}$ into 3 subsets sending the value 0 or 1 or 2. We can now guess the value of G and compute the probability of success following the idea in \cite{Buhrman97}. We shall show that for $k$ large enough all such protocols fail to assure a good guess. For any protocol it is enough to show its failure on some transmission. 

 Start with an example. Suppose there are 10 players. Suppose all players use the protocol $A$ where  $A(\{(0,1),(0,0) \})=0$, $A(\{(1,0),(1,1) \})=1$, and $A(\{(2,0),(2,1) \})=2$. Suppose all players transmit the value 0. So each player reads (0,1) or (0,0) in his/her register. Suppose one of them reads (0,1) and 9 read (0,0).There are 10 such cases where the value of G is 1. Suppose 4 read (0,1) and 6 read (0,0). There are 210 such cases where the value of G is 0. Suppose 7 read (0,1) and 3 read (0,0). There are 120 such cases, where the value of G is 2. And there is the case where they all hold read (0,1) position. For that case the value of G is also 1. All in all there are 341 admissible cases, 11 produce the value 1, 120 the value 2, and 210 the value 0. Therefore if all players use protocol A then it is reasonable to guess that 0 is the value of G. This will be accurate for 210 out of 341 cases.

But it could be that there exists a better protocol. Some of the players might use different divisions. We say that A is a division of type (2,2,2) since each subset of A has 2 elements. There are also divisions of type (3,2,1) and (4,1,1).

The number of possible divisions is finite, so for $k$ large enough one of the above divisions will be used enough times. We shall show that if any division is used enough times then the probability of getting any of the value of G is close to $\frac{1}{3}$. Therefore the best classical protocol completely fails. 

\textbf{ Divisions of type (2,2,2):}
  
Without loss of generality we can assume there are 5 subtypes A,B,C,D,E such that 

\[ A^{-1}(0)= \{(0,0), (0,1) \},\hspace{4mm} B^{-1}(0)= \{(1,0), (0,1) \},\hspace{4mm}   C^{-1}(0)= \{(1,1), (0,1) \}\]

\[D^{-1}(0)= \{(2,0), (0,1) \},\hspace{4mm} E^{-1}(0)= \{(2,1), (0,1) \}\]

\vspace{2mm}

Let (k;1,2,3...) denote $\left(\begin{array}{c}k\\1\end{array}\right)+ \left(\begin{array}{c}k\\2\end{array}\right)+ \left(\begin{array}{c}k\\3\end{array}\right)+...$

If $3j+1$ players are using division A (recall $k=3n+1$ so $j\leq n$) then by counting groups of admissible configurations it is easy to see that the probabilities of getting the values of G are bounded from above by max $\{A^{i,m}\}$  for $i\in\{0,1\}$, $m\in \{0,3,6\}$  where  

\[A^{i,m}=\frac{(3j+1;1+i+m,10+i+m,19+i+m...)}{(3j+1;1+i,4+i,7+i...)}\]

If $j<n$ then all divisions outside $3j+1$ could be of any type. For any state of those outside registers we count all configurations of the $3j+1$ registers of type A. Each such outside state has the same probability. If the number of outside registers with 0 at their bit place is 1 or 0 modulo 3 then the corresponding bound is of type $A^{0,0}$ or $A^{0,3}$ or $A^{0,6}$, otherwise the bound is of type $A^{1,0}$ or $A^{1,3}$ or $A^{1,6}$. 

For $j$ large enough each of the above quotient is close to $\frac{1}{3}$. To see this use Ramus identity \cite{Knuth1968}:

\[\left(\begin{array}{c}n\\q\end{array}\right)+  \left(\begin{array}{c}n\\q+p\end{array}\right)+  \left(\begin{array}{c}n\\q+2p\end{array}\right)+. . . = \frac{1}{p}\sum_{0\leq i<p}{(2cos\frac{i\pi}{p})}^{n} cos\frac{i(n-2q)\pi}{p}.\]

The same argument works for divisions B and D.

The argument for division C or E is similar. In both cases there exists some subset of type $\{(\beta,0), (\alpha,1)\}$, for $\alpha,\beta \in\{0,1,2\}$.

\textbf{ Divisions of type (3,2,1):}

Suppose 3j+1 divisions are of type F where $ F^{-1}(0)= \{(1,0), (1,1), (0,1) \}.$
Note that for any admissible configuration, the sum of the trits of the group of registers with bit value 1 can add up to 0 or 1 or 2, therefore
the probability for any value is bounded from above by:

\[max_{a\in\{0,1,2\}} F^a\]

\noindent where

\[F^a= \frac{ 2\left(\begin{array}{c}3j+1\\a\end{array}\right)+ \sum_{m\in \{3,6,9...\}} \left(\begin{array}{c}3j+1\\a+m\end{array}\right)(a+m;i_m,i_m+3,i_m+6...)}{\sum_{m\in\{0,3,6...\}}\left(\begin{array}{c}3j+1\\a+m\end{array}\right)2^{a+m}}.\]

The variable $a$ depends on the number (modolu 3) of registers with 1 at their bit place outside the set of $3j+1$ of type F. The variable $i_m$ for $m\geq3$ defines a set of sub-configurations of 0 and 1 in the trits of all the registers with 1-value bit, such that their sum (modulo 3) equals $i_m$. So while $m$ is changing, $i_m$ is defined to count all possible sub-configurations with the same value of G. As for $m=0$, to make sure $F^a$ is an upper bound we add the first term in the numerator. 

Note that for $m$ large enough \[\frac{(a+m;i_m,i_m+3,i_m+6...)}{2^{a+m}}\]
\noindent is close to $\frac{1}{3}$ by Ramus identity.

Now split the sums in the numerator and the denominator to a main term and a residue. Then it is clear that for large enough $j$, $F^a$ is close to $\frac{1}{3}$.

In a similar way we can deal with divisions of type H where $ H^{-1}(0)= \{(\gamma,0), (\beta,1), (\alpha,1) \}$, or of type $ I^{-1}(0)= \{(\gamma,1), (\beta,0), (\alpha,0) \}.$ . 

The case for divisios of type J where $ J^{-1}(0)= \{(\gamma,1), (\beta,1), (\alpha,1) \}$, or of type K where $ K^{-1}(0)= \{(\gamma,0), (\beta,0), (\alpha,0) \}$ is trivial.

\textbf{ Divisions of type (4,1,1):}

Consider divisions of type L, where $ L^{-1}(0)= \{(1,0), (0,0), (1,1), (0,1) \}$. Each configuration has $m$ registers with 0 value bit and $3j+1-m$ registers with 1 value bit. Each such configuration has $2^{3j+1}$ sub-configurations if we include all the states the trits could have. All sub-configuration can be grouped as follows. Given $i_m$ in $\{0,1,2\}$ a function of $m$, we can group all 1-value bit registers (if there are any) having $b$ as the sum of all their trits with the 0-value bit registers (if there are any) having $c$ as the sum of their trits, this we do for all $b$ and $c$ such that $b+c=i_m$ modulo 3. So while $m$ is changing, $i_m$ is defined to count all possible sub-configurations with the same value of G.

Thus the probability of success is bounded from above by:

\[max_{a\in\{0,1,2\}} L^a\]

\noindent where

\[ L^a= \frac{ \sum_{m\in\{0,3,6...\}} \left(\begin{array}{c}3j+1\\a+m\end{array}\right)\sum_{b+c =i_m} (a+m;b,b+3,...)'  (3j+1-(a+m);c,c+3,...)' }{ \sum_{m\in\{0,3,6...\}} \left(\begin{array}{c}3j+1\\a+m\end{array}\right)2^{3j+1}}\]

\noindent and where $ (k;1,2,3...)' =(k;1,2,3...)$ for $k>0$, otherwise $ (k;1,2,3...)' =1$.

For $m$ big enough each one of the terms in the inner sum is close to $\frac{1}{9}$, there are three such terms so the inner sum is close to $\frac{1}{3}$.

Now split the exterior sum in the numerator and the sum in the denominator, each into three terms, a main term and two residues (for $m$ too big or too small). Then we can use simple properties of binomial coefficients to conclude.

In a similar way it is possible to treat the case $ M^{-1}(0)= \{(\delta,0), (\gamma,0), (\beta,1), (\alpha,1) \}$.

Consider divisions of type N, where $ N^{-1}(0)= \{\alpha,0), (2,1), (1,1), (0,1) \}$.
Then the probability is bounded from above by:

\[max_{a\in\{0,1,2\}} N^a\]
\noindent where
\[ N^a= \frac{\sum_{m\in\{0,3,6...\}} \left(\begin{array}{c}3j+1\\m+a\end{array}\right)3^{ +(m+a-1) }}{\sum_{m\in\{0,3,6...\}} \left(\begin{array}{c}3j+1\\m+a\end{array}\right)3^{m+a}}\]

\noindent where $+(z)=z$ if $z$ is a positive integer, otherwise $+(z)=0.$

Clearly this quotient goes very fast to $\frac{1}{3}$.  

In a similar way we can deal with divisions of type O where $ O^{-1}(0)= \{(\alpha,1), (2,0),  (1,0), (0,0) \}$.


\bibliographystyle{unsrt}

\def\searchbib#1{\IfFileExists{#1.bib}{\bibliography{#1}\end{document}}{}}
\end{document}